\documentclass{article}
\usepackage{graphicx}

\usepackage{amsmath}
\usepackage{amssymb}

\usepackage{bbm}
\newcommand{\M}{M_{\rm pl}}

\DeclareFontFamily{OT1}{rsfs}{}
\DeclareFontShape{OT1}{rsfs}{m}{n}{ <-7> rsfs5 <7-10> rsfs7 <10->rsfs10}{} 
\DeclareMathAlphabet{\mycal}{OT1}{rsfs}{m}{n}
\newcommand{\lan}{{\mycal L}}

\newcommand{\dual}[1]{\overset{{}^{{}^{\boldsymbol{\neg}}}}{\smash[t]{#1}}}

\begin{document}

\title{Dark energy with dark spinors}

\author{Christian G. B\"ohmer\footnote{\texttt{c.boehmer@ucl.ac.uk}}\mbox{\ } and James Burnett\footnote{\texttt{j.burnett@ucl.ac.uk}}\\ Department of Mathematics, University College London,\\ Gower Street, London, WC1E 6BT, UK}

\date{\today}

\maketitle

\begin{abstract}
Ever since the first observations that we are living in an accelerating universe, it has been asked what dark energy is. There are various explanations all of which with have various draw backs or inconsistencies. Here we show that using a dark spinor field it is possible to have an equation of state that crosses the phantom divide, becoming a dark phantom spinor which evolves into dark energy. This type of equation of state has been mildly favored by experimental data, however, in the past there were hardly any candidate theories that satisfied this crossing without creating ghosts or causing a singularity which results in the universe essentially ripping. The dark spinor model converges to dark energy in a reasonable time frame avoiding the big rip and without attaining negative kinetic energy as it crosses the phantom divide.
\end{abstract}


\section{Introduction}

An increasing number of independent observations indicates that we are living in an expanding universe where the expansion itself is accelerating~\cite{Spergel:2006hy,Tegmark:2006az,Percival:2007yw}. It has been accepted that this requires some additional negative-pressure matter source, named dark energy. The simplest model explaining this accelerated expansion is the cosmological constant $\Lambda$ which corresponds to an unusual equation of state $w=P/\rho=-1$. Although the standard model of cosmology, the $\Lambda$ cold dark matter ($\Lambda$CDM) model, fits the present data very well, the numerical value of the cosmological constant is about 120 orders of magnitude smaller than the vacuum expectation value predicted by quantum field theory. This smallness problem can be addressed by considering dynamical models, where the field slowly rolls down some potential where the effective equation of state $w_{\rm eff}$ converges to $w_{\rm eff}=-1$. In the past it has been believed that this value should be approached from above. However, there has been a recent interest in phantom models where the dark energy equation of state is approached from below $w \leq -1$, see~\cite{Caldwell:1999ew,Caldwell:2003vq,Singh:2003vx,Nojiri:2003vn,Gibbons:2003yj,GonzalezDiaz:2003bc,Dabrowski:2003jm,Chimento:2003qy,Stefancic:2003rc,Johri:2003rh,Vikman:2004dc,Hu:2004kh,Stefancic:2003bj,Sola:2005et,Capozziello:2005tf,Nesseris:2006er}. These models, as counter intuitive as they may appear, are not excluded by current data~\cite{Caldwell:1999ew,Caldwell:2003vq}.

\begin{figure}[!t]
\centering
\includegraphics[scale = 0.4]{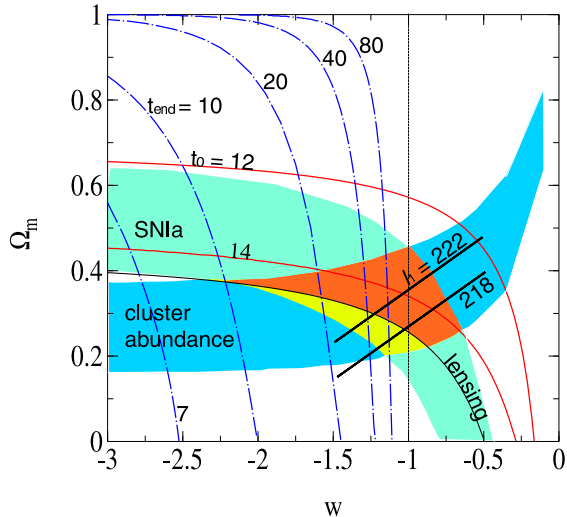}
\caption{Current constraints to the $w-\Omega_m$ parameter space. The red solid curves show the age (in Gyr) of the Universe today (assuming a Hubble parameter $H_0 =70\,\mathrm{km}\,\mathrm{sec}^{-1}\,\mathrm{Mpc}^{-1}$). The light shaded regions are those allowed (at 2$\sigma$ confidence level) by the observed cluster abundance and by current supernova measurements of the expansion history. The dark orange shaded region shows the intersection of the cluster-abundance and supernova curves, additionally restricted (at 2$\sigma$ confidence level) by the location of the first acoustic peak in the cosmic-microwave-background power spectrum and quasar-lensing statistics.}
\label{fig1}
\end{figure}

Figure~\ref{fig1}, taken from~\cite{Caldwell:2003vq}, shows data taken from the cluster abundance, supernovae, quasar-lensing statistics and the first acoustic peak in the cosmic microwave background radiation (CMB) power spectrum, which all imply that there is convergence to dark energy. However, when the parameter space is expanded to include the phantom region the data does not rule out a convergence from below~\cite{Caldwell:2003vq}. In fact, the data even seems to indicate that this is quite likely.

Considering such phantom energy models, the outcome of such a universe dominated by phantom energy is very different to any outcome we are accustomed to. The evolution parameter increases quicker than the horizon and quickly starts to pull apart gravitationally bound objects and then finally any bound objects, including the strong force! However, since cosmologists turned their attention from a constant equation of state for dark energy to one where it evolves through time it is believed that with or without crossing the divide the equation of state should eventually converge to $w=-1$ due to the success of the $\Lambda$CDM model, and that observationally it is so close to this value at present. The majority of such dynamical dark energy models is based on evolving scalar fields with a suitably chosen potential. However, a canonical scalar field cannot cross the phantom divide and therefore such models have to modified, for instance by considering fields with negative kinetic energy.

Within theories of modified gravity one is left with the option of either changing the geometrical side or the matter side of the gravitational field equations, changing either gravity itself or altering the matter content which contributes to the forces, respectively. Those two approaches are not entirely independent as many modified theories bring the new geometrical quantities to the matter side, ultimately changing the energy content of the universe, however allowing for a new interpretation.

Here, we follow a different approach, namely we model the evolving dark energy component of the universe to be of the form of a single dark spinor field. Dark spinors, originally dubbed the Elko spinor~\cite{jcap}, are similar to Majorana spinors but acquire the full four degrees of freedom of a Dirac spinor due to their helicity structure. They couple to the Higgs mechanism via $\dual{\lambda}\lambda H^{\dagger}H$ and weakly to the electromagnetic field via $\dual{\lambda}[\gamma^a,\gamma^b]\lambda F_{ab}$, however in the latter case this coupling is heavily constrained because of the masslessness of the photon, making them a candidate for dark matter. The idea of one field explaining both dark matter and dark energy has already been discussed in various approaches, see e.g.~\cite{Boyle:2001du,Capozziello:2005tf}. One possible mass range for our dark spinors is in the $m \simeq \mathrm{MeV}$ range. 

The dark spinor field is based on the eigenspinors of the charge conjugation operator, for a general introduction we refer the reader to~\cite{jcap}. For the subsequent developments of dark spinors as an alternative model to scalar field inflation, see~\cite{Boehmer:2006qq,Boehmer:2007dh,Boehmer:2008ah}. For instance, in~\cite{Boehmer:2007ut} it could be shown that dark spinors have the effect of suppressing the low multipole amplitude of the primordial power spectrum, while at the same time to provide the usual inflationary features and predicting a small scale power spectrum in agreement with scalar field inflation, see in particular~\cite{Boehmer:2008rz,Gredat:2008qf,Shankaranarayanan:2009sz} for dark spinors in cosmology.

\section{Cosmological dark spinor field equations}

The standard model of cosmology is based upon the flat Friedmann-Lema\^{\i}tre-Robertson-Walker (FLRW) metric
\begin{align}
  ds^2 = dt^2 - a(t)^2 (dx^2+dy^2+dz^2),
  \label{flrw}
\end{align}
where $a(t)$ is the scale factor and $t$ is cosmological time. The dynamical behavior of the universe is determined by the cosmological field equations of general relativity
\begin{align}
  R_{\alpha\beta} - \frac{1}{2}R g_{\alpha\beta} = \frac{1}{\M^2} T_{\alpha\beta},
  \label{eq1}
\end{align}
where $\M$ is the Planck mass which we use as the coupling constant, $1/\M^2=8\pi G$ and $c=1$. $T_{\alpha\beta}$ denotes the stress-energy tensor, which for a homogeneous and isotropic cosmology takes the form
\begin{align}
  T_{\alpha\beta} = \mbox{diag}(\rho,a^2 P,a^2 P,a^2 P).
  \label{eq2}
\end{align}
The cosmological field equations are given by
\begin{align}
  H^2 = \frac{1}{3\M^2}\rho,
  \label{eq5}\\
  \dot{\rho} + 3H(\rho + P) = 0.
  \label{eq7}
\end{align}
where the dot denotes differentiation with respect to time $t$ and the Hubble parameter $H$ is defined by $H=\dot{a}/a$.

Let us consider a homogeneous single dark spinor field. Following~\cite{Boehmer:2008rz,Gredat:2008qf}, the effective Lagrangian density of this field can be written in terms of the {\em scalar} field $\varphi$ as
\begin{align}
  \lan = \frac{1}{2} \dot{\varphi}^2  + \frac{3}{8}H^2 \varphi^2 - V(\varphi).
  \label{eq:n1}
\end{align}
If the potential $V(\varphi)$ contains a standard mass term $V(\varphi)=m^2 \varphi^2/2$, then we can rewrite the Lagrangian as
\begin{align}
  \lan = \frac{1}{2} \dot{\varphi}^2  + \frac{3}{8}H^2 \varphi^2 - \frac{1}{2} m^2 \varphi^2.
  \label{eq:n2}
\end{align}
This allows us to interpret the explicit presence of the Hubble parameter in the action as an effective mass term where the mass changes as the universe evolves, and we have
\begin{align}
  m_{\rm eff}^2 = m^2 - \frac{3}{4} H^2.
  \label{eq:n3}
\end{align}
If the universe undergoes a phase of accelerated expansion, the Hubble parameter is approximately constant. Depending on the ratio $m/H$, it is possible for models to attain a negative value for $m_{\rm eff}^2$ without creating ghost of having negative kinetic energy. This is possible because of the additional coupling of geometry to the matter field we are considering, a spinor opposed to a scalar field. 

The energy density and the pressure of the dark spinor field are given by
\begin{align}
  \rho_\varphi &= \frac{1}{2}\dot{\varphi}^2 + V(\varphi) - \frac{3}{8}H^2\varphi^2,
  \label{eq8}\\
  P_\varphi &= \frac{1}{2}\dot{\varphi}^2 - V(\varphi) + \frac{1}{8}H^2\varphi^2.
  \label{eq9}
\end{align}
These two equations have the important property of leaving the acceleration equation unchanged
\begin{align}
  \frac{\ddot{a}}{a} = - \frac{1}{3\M}(\dot{\varphi}^2-V(\varphi)).
\end{align}
The spinor field's potential energy may yield an accelerated expansion of the universe. It should be noted that the energy density and the pressure now explicitely depend on the Hubble parameter. These additional terms are present because the covariant derivative when acting on the spinor field also acts on its spinorial part. One way of interpreting this `coupling' in Eq.~(\ref{eq:n1}) is to regard the effective mass of the particle to depend on the Hubble parameter~\cite{Boehmer:2008rz} and therefore on the evolution of the universe. Another interesting way of reading this equation is to regard the gravitational coupling to be time dependent~\cite{Gredat:2008qf}.

The effective equation of state of the dark spinor field is given by
\begin{align}
  w_{\rm eff} = \frac{P_\varphi}{\rho_\varphi} = \frac{\frac{1}{2}\dot{\varphi}^2 + V(\varphi) - \frac{3}{8}H^2\varphi^2}{\frac{1}{2}\dot{\varphi}^2 - V(\varphi) + \frac{1}{8}H^2\varphi^2},
\end{align}
and its more complicated form, when compared with the scalar field, also shows that a crossing of the phantom divide is possible without negative kinetic energy. If we for the moment take the canonical mass term for the potential and assume that $\dot{\varphi}^2 \ll V(\varphi)$, the effective equation of state becomes
\begin{align}
  w_{\rm eff} \simeq -\frac{m^2 - \frac{3}{4}H^2}{m^2 - \frac{1}{4}H^2} = 
  -\frac{1 - \frac{3}{4}H^2/m^2}{1 - \frac{1}{4}H^2/m^2} \geq -4.
\end{align}
Therefore, we expect dark spinor models with canonical mass term to cross the phantom divide, however, we do not expect the effective equation of state to diverge since once the field is slow-rolling in its potential, there exists a lower bound to $w_{\rm eff}$.

While there is a large number of potentials considered in the literature in the context of dynamical dark energy models, we will restrict our attention to power counting renormalizable potentials. As the dark spinor field has mass dimension one, the two allowed potentials are
\begin{align}
  V_1(\varphi) = \frac{1}{2} m^2 \varphi^2,
\end{align}
and
\begin{align}
  V_2(\varphi) =  \frac{1}{2} m^2 \varphi^2 + \frac{1}{4} \alpha \varphi^4,
\end{align}
where the $V_1(\varphi)$ is the aforementioned canonical mass term and $V_2(\varphi)$ includes the self interaction term with $\alpha$ being a dimensionless coupling constant. 

\section{Dark spinors as dark energy}

Let us start by solving Eq.~(\ref{eq5}) for the Hubble parameter, using Eq.~(\ref{eq8}) we find
\begin{align}
  H = \frac{1}{\sqrt{3}\M}\frac{\sqrt{\dot{\varphi}^2/2 + V(\varphi)}}
  {\sqrt{1 + (\varphi/\M)^2/8}}.
  \label{eq:n4}
\end{align}
Hence, using the first equation, the energy density of the dark spinor field is
\begin{align}
  \rho_\varphi &= \frac{1}{2}\dot{\varphi}^2 + V(\varphi) - 
  \frac{1}{8}\frac{\dot{\varphi}^2/2 + V(\varphi)}{1 + (\varphi/\M)^2/8}
  (\varphi/\M)^2 \\[1ex]
  &= \Bigl(\frac{1}{2}\dot{\varphi}^2 + V(\varphi)\Bigr)\Bigl(1-\frac{(\varphi/\M)^2/8}{1 + (\varphi/\M)^2/8}\Bigr).
  \label{eq:n5}
\end{align}
It is precisely this latter form of the energy density which motivated~\cite{Gredat:2008qf} to interpret~(\ref{eq:n5}) as inducing a time-dependent gravitational coupling by considering $G_{tt} = 8\pi G_{\rm eff} \bar{\rho}_\varphi$ where $\bar{\rho}_\varphi$ is the standard energy density of a scalar field.

\subsection{Phantom dark energy models}

Now, we consider the conservation equation~(\ref{eq7}) with~(\ref{eq:n4}) and~(\ref{eq:n5}) and numerically solve the resulting equation for $\varphi(t)$ and substitute into Eqs.~(\ref{eq8}) and~(\ref{eq9}) to obtain the evolution of the effective equation of state $w_{\rm eff} = P/\rho$ and plot it as a function of the evolution parameter $a(t)$. We chose our initial conditions to be $w(0)=\{ 1/3, 0, -1/3, -2/3 \}$, the first two representing radiation and dust, respectively. The initial conditions with $w(0) \leq -1/3$ correspond to an initially accelerating universe. Small changes in these initial conditions do not alter the late-time asymptotic behavior of the solutions.

\begin{figure}[!tb]
\centering
\includegraphics[scale = 0.8]{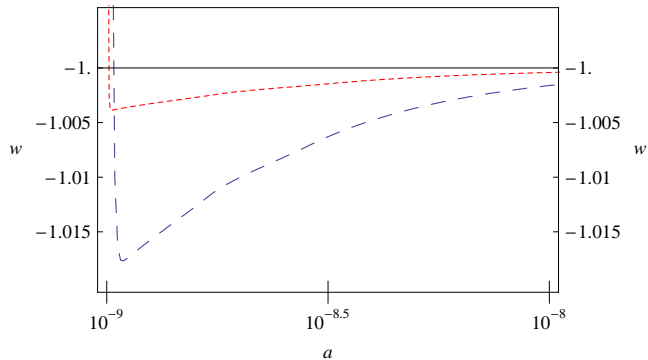}
\caption{Equation of state for $V_1(\varphi)$: $\M=1$, $\dot\varphi(0)=1$  and $w(0)=-1/3$. With $m^2 = \{0.002,0.001\}=\{\mathrm{red\ (higher)},\mathrm{blue\ (lower)}\}$ }
\label{fig1w}
\end{figure}

\begin{figure}[!tb]
\centering
\includegraphics[scale = 0.8]{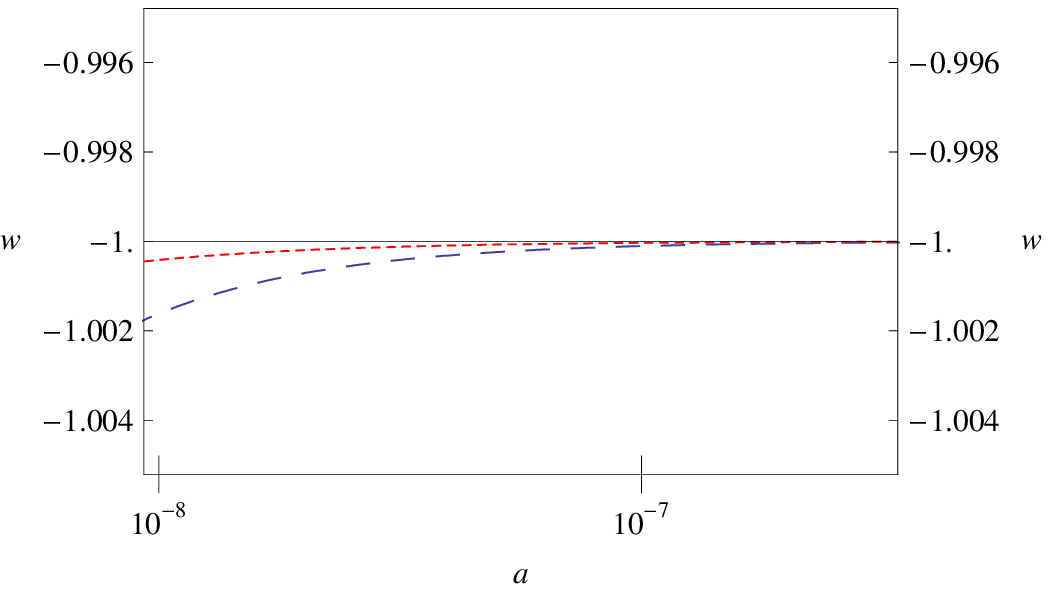}
\caption{Equation of state for $V_1(\varphi)$: $\M = 1$, $\dot\varphi(0)=1$ and $w(0)=-1/3$. With $m^2 = \{0.002,0.001\}=\{\mathrm{red\ (higher)},\mathrm{blue\ (lower)}\}$.}
\label{fig2w}
\end{figure}

Fig.~\ref{fig1w} shows the dynamical behavior of the effective equation of state considering the potential $V_1$ with $\M = 1$, $\dot\varphi(0)=1$ and $w(0)=-1/3$. For the two different mass values it is possible to see that the effective equation of state almost immediately drops below the phantom divide. During the subsequent evolution, $w$ begins to increase as further shown by Fig.~\ref{fig2w} to the desired dark energy value. From Fig.~\ref{fig2w} it is also evident that our model is practically indistinguishable from dark energy modeled by a cosmological constant, long before recombination when the scale factor is about $a(t)=10^{-3}$.

\begin{figure}[!tb]
\centering
\includegraphics[scale = 0.8]{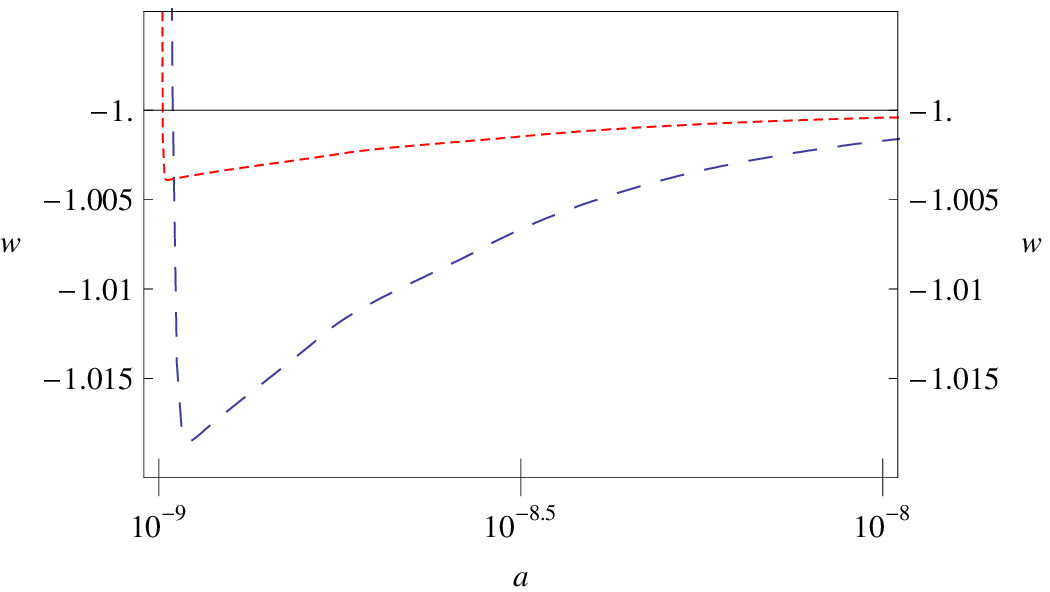}
\caption{Equation of state for $V_1(\varphi)$: $\M = 1$, $\dot\varphi(0)=1$ and $w(0)=1/3$. With $m^2 = \{0.002,0.001\}=\{\mathrm{red\ (higher)},\mathrm{blue\ (lower)}\}$.}
\label{figwintsamemass}
\end{figure}

\begin{figure}[!tb]
\centering
\includegraphics[scale = 0.8]{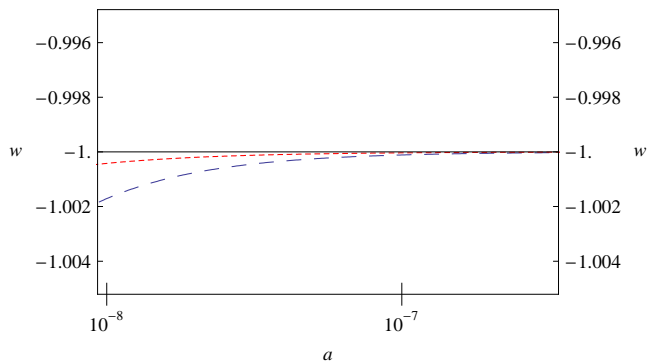}
\caption{Equation of state for $V_1(\varphi)$: $\M = 1$, $\dot\varphi(0)=1$ and $w(0)=1/3$. With $m^2 = \{0.002,0.001\}=\{\mathrm{red\ (higher)},\mathrm{blue\ (lower)}\}$.}
\label{fig2wint}
\end{figure}

We obtained very similar results for other initial values of the equation of state: $w(0)=1/3$, $w(0)=0$ and $w(0)=-2/3$. We have shown, for comparison, results from $w(0)=1/3$ in Fig.~\ref{figwintsamemass} and Fig.~\ref{fig2wint}, they qualitatively agree with the results presented in Figs.~\ref{fig1w} and~\ref{fig2w}, respectively.

\subsection{Diverging models}

The next natural step is to consider a potential which allows for a self-interaction terms. As discussed above, the inclusion of this term also yields a power counting renormalizable field theory.

\begin{figure}[!tb]
\centering
\includegraphics[scale = 0.75]{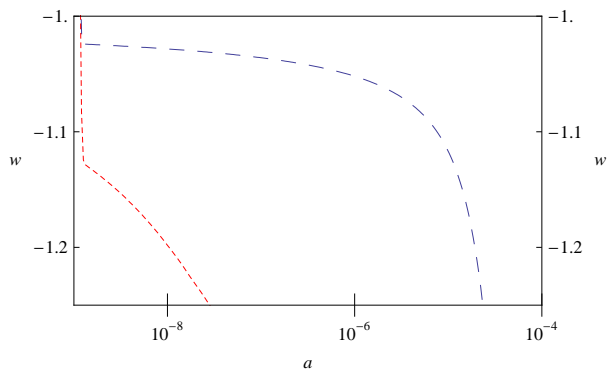}
\caption{Equation of state for $V_2(\varphi)$: $\M = 1$, $w(0)=1/3$ and $\alpha=1$. With $m^2 = \{4,0.02\}=\{\mathrm{red\ (higher)},\mathrm{blue\ (lower)}\}$.}
\label{crazy}
\end{figure}

When we include the self interaction term in the potential, $V_2(\varphi)$, we find, for all initial values of $w$ and $\alpha = 1$, that the effective equation of state always diverges to $-\infty$, see Fig.~(\ref{crazy}). We have also checked, although not included in here, that our numerical results for $V_2(\varphi)$ converge to results for $V_1(\varphi)$ as $\alpha \rightarrow 0$. Although it might be possible to construct models with finely tuned initial conditions such that the divergence of the equation of state would happen in the future, such models are very unlikely. Hence, we are lead to conclude that a dynamical dark energy model based on dark spinors requires their potential to be of the simplest form, namely a canonical mass term, without self interaction.

\subsection{Oscillating models}

\begin{figure}[!htb]
\centering
\includegraphics[scale = 0.75]{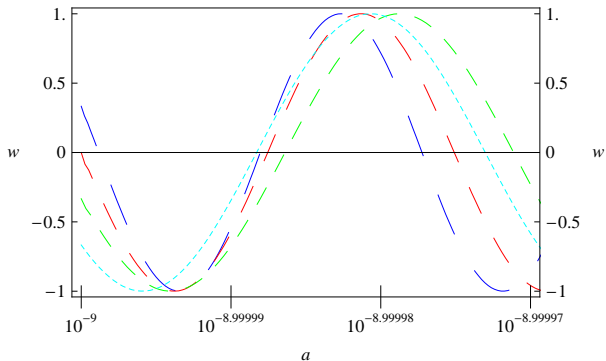}
\caption{Equation of state for potential $V_1(\varphi)$: $\M = 1$, $m=0.1$ with initial conditions chosen such that $w(0)=\{1/3,0,-1/3,-2/3\}$, respectively $\{\mathrm{blue (long dashed},\mathrm{red\ (medium\ dashed)},\mathrm{green\ (dashed)},\mathrm{cyan\ (short\ dashed)}\}$}
\label{crazy1}
\end{figure}

Lastly, we found another set of interesting results where the equation of state oscillated between $w=1$ and $w=-1$ for all time. The oscillation of the equation of state is very rapid, as can be seen in Fig.~\ref{crazy1}, making them unphysical as a dark energy model. This qualitative behavior does not change if we include the self-interaction term. However, if such a model could be modified it would be a prime candidate for models where the field changes its characteristic from being dark matter at early times to become dark energy at late times, see also~\cite{Cen:2000xv,Oguri:2003nn,Malik:2002jb,Ziaeepour:2003qs,Boehmer:2008av}.

\section{Conclusions and discussions}

A dark spinor field is able to provide a possible model for dark matter as it couples mainly via the Higgs mechanism, but has heavily constrained interactions with the electromagnetic field. Dark spinors have a predicted $\mathrm{MeV}$ mass range and therefore experimental predictions can be formulated and possibly measured at the LHC. Our results now show that the dark spinor field is also capable of having a dynamical equation of state which crosses the phantom divide and asymptotes to $w=-1$. This makes it a viable candidate for dark energy which cannot be ruled out experimentally. 

Unlike previous phantom models, dark spinors do not obtain negative kinetic energy on crossing the phantom divide, due to both $\rho$ and $P$ depending on the Hubble parameter and therefore these models do not create ghosts. According to~\cite{Caldwell:2003vq} the equation of state must not stay below the divide but converge to dark energy, therefore the dark spinors' potential is of the simplest form, a canonical mass term $m^2 \varphi^2/2$. Our dark spinor model does not require a modification of general relativity, leaving one of the most successful models in theoretical physics untouched.

Due to the interesting nature of dark spinors, they have been shown to give other unique properties not found with other matter sources considered in the past. For now, in a cosmological setting, dark spinors are providing intriguing results in having the potential to be the best candidate dynamical dark energy model at hand. 

\subsection*{Acknowledgments}
We thank Jochen Weller for useful discussions and comments on the manuscript.

\end{document}